\documentclass[12pt,onecolumn,pagenumbers,noSBCmetadata,brazilian,english]{sbc20}

\usepackage{verbatim}  
\usepackage{graphicx}  
\usepackage{array}     
\usepackage{mathtools} 

\usepackage{booktabs}  

\usepackage{indentfirst}

\usepackage{pifont} 

\publicationInfo
  {
    name={Institute of Mathematics and Statistics, University of São Paulo \\
          \textbf{Technical Report (written in Aug 2019) -- RT-MAC-2022-01}},
    copyrightyear=2022,
    bibstyle=sbc20,
    nousecategory,
    license=CC-BY-NC-ND,
  }

\articleInfo
  {
  }

\title{The Next-Generation OS Process Abstraction}

\author
  [
    email=siqueirajordao@riseup.net,
    orcid=0000-0001-6879-1978,
    institutionID=amd,
    country=Canada,
  ]
  {Rodrigo Siqueira}

\author
  [
    email=lago@ime.usp.br,
    orcid=0000-0002-4306-8078,
    institutionID={usp},
    country=Brazil,
  ]
  {Nelson Lago}

\author
  [
    email=fabio.kon@ime.usp.br,
    orcid=0000-0003-3888-7340,
    institutionID=usp,
    country=Brazil,
  ]
  {Fabio Kon}

\author
  [
    email=dejan.milojicic@hpe.com,
    orcid=0000-0001-9830-8588,
    institutionID=hp,
    country=USA,
  ]
  {Dejan Milojičić}

\institution{amd}{Advanced Micro Devices}
\institution{usp}
  {Institute of Mathematics and Statistics \sep University of São Paulo}

\institution{hp}
  {Hewlett Packard Labs}

\shortauthor{Siqueira et al.}


\abstract{
 Operating Systems are built upon a set of abstractions to provide resource
 management and programming APIs for common functionality, such as
 synchronization, communication, protection, and I/O. The process abstraction
 is the bridge across these two aspects; unsurprisingly, research efforts pay
 particular attention to the process abstraction, aiming at enhancing security,
 improving performance, and supporting hardware innovations. However, given the
 intrinsic difficulties to implement modifications at the OS level, recent
 endeavors have not yet been widely adopted in production-oriented OSes. Still,
 we believe current hardware evolution and new application requirements provide
 favorable conditions to change this trend. This paper evaluates recent
 research on OS process features identifying potential evolution paths. We
 derive a set of relevant process characteristics, and propose how to extend
 them as to benefit OSes and applications.
}

\begin{document}

\maketitle

Modern OS have a dual role: on the one hand, they provide a set of abstractions
built on top of hardware devices to offer features for user applications; on
the other hand, they offer fundamental programming APIs for common
functionality, such as synchronization, communication, protection, and I/O.
Given these two central roles, it is not surprising that small improvements in
OSes, including non-functional aspects such as performance, fault tolerance,
security and isolation, may result in significant benefits for a large number
of applications. Accordingly, there is significant pressure for OSes to address
both software and hardware evolution. New application areas such as
Machine/Deep Learning, Microservices, and Smart Cities require faster remote
data accesses, new abstraction layers, reduced complexity to use resources, and
security improvements. At the same time, emerging hardware trends such as the
move from the current homogeneous and CPU-centric to a heterogeneous and
Memory-centric computing architecture, Non-volatile Memories, and FPGAs
(Field-Programmable Gate Arrays) show great promise, but only inasmuch as they
are properly supported by the OS.

In most of the current production-oriented OSes (e.g., GNU/Linux Distributions,
Windows, MacOS, and others), the process abstraction is the meeting point of
the hardware abstractions, the OS API entry points, and the (often multiple)
user applications. Virtually all hardware access and OS services occur in
response to process requests; at the same time, security, isolation, etc. are
applied almost entirely to processes. Therefore, it is a perfect target for
improvements in OS design.\looseness=-1

Despite their potential benefits, however, proposals of changes in the
established OS mechanisms are usually met with skepticism because, by the same
measure, they may bring about instabilities, security vulnerabilities,
fragility, and backward compatibility issues. To make matters worse, many of
the current proposals of changes to the process abstraction are not yet mature,
lacking robust implementations, and the ideas require refinements.
Consequently, user applications create sophisticated techniques to circumvent
problems that could be simplified by updating the OS abstractions.

Over the years, a vast amount of research has been carried out in an attempt to
expand the process abstraction and exploit new features. Given the new hardware
features and emerging applications that shall dominate the computing field in
the next few years, we believe it is time to rethink how processes concepts
work and review these proposals to find a tradeoff between the state-of-the-art
of research proposals and the state-of-the-practice. As the first step to this
end, this article categorizes some of the main properties of the process
abstraction, examines previous works that propose extensions, discuss potential
for adoption, and present an outlook for the field.

\subsection*{New Hardware Features}

Usually, changes in hardware are complicated and take a long time to develop,
implement, and deploy to the market. Still, advances in technology and new
requirements are currently reshaping the hardware landscape, enabling adoption
of significant improvements in chip design. Given this fluidity, hardware
designers have the chance to revisit the vast research in the area for sources
of inspiration.

Recent hardware vulnerabilities, such as Meltdown and Spectre, have pushed the
industry to review their processor design; it is convenient to inspect
proposals for additional hardware mechanisms that improve isolation and memory
translation, either by fine-grained memory protection or virtualization
mechanisms. Technologies such as heterogeneous accelerators and the RISC-V
project open new venues for innovation; industry and academia alike have been
conducting studies on accelerator devices (GPUs, FPGAs, ASICs, DSPs, etc.) and
how to effectively deploy heterogeneous computing systems to users for improved
performance and flexibility. Finally, emerging technologies have the potential
to drive the adoption of new programming paradigms. Thanks to NVM, we expect
that in a few years computers will be able to access petabytes of data by using
rack-scale systems; such availability of high-density NVMs shall change the way
OSes are built.\looseness=-1

\subsection*{Emerging Applications}

Newer areas of application, such as machine \& deep learning, microservices,
big data, IoT, and smart cities platforms, differ significantly from more
traditional applications in their need for (1) storage of vast amounts of data,
(2) efficient local and wide area data sharing and communications, (3) high
processing throughput, and (4) power efficiency. These new demands may benefit
from several OS improvements and new programming models that enrich the palette
of programming patterns available to developers.

It is these new programming models that open countless opportunities for
programmers, who use their creativity to solve problems in new and unimagined
ways. For example, when threads were first released just a few techniques
existed but, after a few years, a vast number of threading patterns became
widespread.

One clear case of the demand for such new paradigms is the trend towards
microservices. Microservices are a modern approach to bring modularization,
fault-tolerance, and scalability to large-scale systems. Many
microservices-based applications, such as Smart Cities platforms and
applications, require the intense use of Cloud Computing, Big Data, and IoT
technologies to provide persistent and real-time data. Their advantages,
however, are coupled with integration, storage, and data sharing challenges,
increases in system complexity and communications overhead, and complex scaling
mechanisms. Accordingly, better support for new and legacy modularized
applications may bring immediate benefits for microservices.

New programming models also represent opportunities to renew user applications
and continuously optimize and improve them. Since there is also a vast amount
of legacy applications that could be reused in these new environments, provided
they can incorporate security improvements, better modularization mechanisms,
runtime updating capabilities, and code simplification, mechanisms that allow
new paradigms to be easily incorporated into legacy code offer excellent
opportunities for code reuse.

\section{Process Targeted Aspects}

We believe that the process abstraction represents one of the main entry points
for bringing innovations that address the new demands imposed by new hardware
and software trends. In this section, we highlight the most visible parts of
the process abstraction that have been the target of recent explorations.

\subsection{Programming Models}

Besides the manipulation of hardware devices, OSes provide several additional
features to user applications, such as file locking and security primitives. To
use any of them, the application should be able to access it by means of a
coherent programming model, i.e., a set of well established interrelated
abstractions. A given OS implements such programming models into its own,
specific APIs~\cite{cardoso:2011}. For example, both GNU/Linux and Windows
provide different threading APIs (\textit{pthreads} and \textit{WindowsThreads}),
but both correspond to the same parallel programming model.

Currently, most OSes support several widely used programming models and their
corresponding APIs. Nevertheless, user applications have been changing over the
years, and demands for improvements in areas such as better security layers,
optimization options, and code simplification are a real issue. In this sense,
proposals for expanding the process abstraction by means of new programming
models represent an interesting innovation path to support modern user space
applications.

The execution flow of a process is controlled by its Program Counter (PC), and
the OS Scheduler is responsible for retrieving its execution context before the
next instructions, as indicated by the PC, can be executed. Accordingly, both
the PC and the rest of the kernel level data about the process (memory map,
file descriptor table, etc.) are kept together and manipulated by the Scheduler
as a single entity (the \textit{task}).

Departing from this idea, Litton et al. propose to decouple the PC and the
Scheduler from the rest of the process context data: with Lightweight Contexts
(\textit{lw}C), a single process (with a single PC) may have multiple different
internal contexts~\cite{lwc:2016}. This allows us to perform interesting manipulations,
such as creating a snapshot of the current process state and, later on,
reverting to this previous state. Another possibility is switching to a new
process context with restricted access to memory regions before the execution
of security-sensitive code. This is not unlike using independent processes or
threads but, instead of relying on the Scheduler to swap the process context
(which is reasonably expensive), it is the application itself that chooses when
and how to manipulate the context, bringing finer control to the programmer
along with better performance.

An \textit{lw}C comprises a virtual memory mapping, a collection of page
mappings, file descriptor bindings, and a set of credentials; whenever a new
process is created, the system creates a new \textit{lw}C for it. The
application may access all \textit{lw}C features directly from user space
through system calls that fine-tune its behavior. The most important ones are
\texttt{lwCreate} and \texttt{lwSwitch}, which have semantics similar to
\texttt{fork}: after \texttt{lwCreate} returns, the current process has a new
\textit{lw}C (child) associated with itself which is a snapshot of the caller
process. This snapshot differs from fork because no new PID or thread is
created (because these only make sense to track different PCs). After the child
is created, the application is free to switch (\texttt{lwSwitch}) back to the
snapshot at any given time.\looseness=-1

Most of the OSes impose few security restrictions to processes by default.
Process creation illustrates this argument: when a parent process creates a
child, all of its data reflects the parent data. Whenever a programmer wants to
restrict permissions, they must spend considerable effort due to the permissive
defaults of current OSes. Bittau et al. proposed a new approach named
Wedge~\cite{wedge:2008} to address these issues; they introduced a model wherein the
OS provides primitives that create compartments with default-deny semantics.

While applications could be compartmentalized without OS support, most follow a
monolithic design, with no clear separation between elements, because it is
much simpler to do so. Wedge improves this scenario with three primitives:
\textit{tagged memory}, \textit{callgates}, and \textit{sthreads}. Tagged
memory is a mechanism to declare memory access privileges: the programmer
creates a new tag (e.g., \texttt{t=read-write}) and allocates memory (with the
\texttt{smalloc} system call) using it as a parameter. Callgates are
responsible for executing code with different privileges on behalf of the
caller. Sthreads are the central component of Wedge, responsible for providing
isolation units. They are composed of a control thread and a security policy,
which specifies information such as memory tags and permissions, file
descriptor access, and associated callgates.

The programming model used by Wedge enables programmers to easily
compartmentalize legacy applications using a simple set of operations in
well-defined places. The authors also created a tool named Crowbar to support
developers in the use of the Wedge primitives by analyzing running code and
identifying potential places to create compartments. The authors modified
Apache/OpenSSL and OpenSSH using this approach and showed that many well-known
vulnerabilities became ineffective.

Some researchers pointed to other problems associated with the process
abstraction: the representation of a pointer-based data structure outside the
process limits, the annoyance associated with the task of coordinating shared
memory access by multiple processes, and the problem of addressing high-density
physical memory. While there are solutions for all of these problems, Hajj et
al. argued for a new approach named SpaceJMP~\cite{spacejmp:2016} that may solve
part of them. Traditionally, processes have just one associated Virtual Address
Space (VAS); in contrast, SpaceJMP can detach VASes from processes, enabling a
single process to have multiple VASes -- and, therefore, multiple execution
contexts.

Consider a simple process that starts with a default VAS associated with it, as
happens with traditional processes. After the programmer invokes the
\texttt{vas\_create} system call, a new VAS is created. Given such an existing
VAS, the function \texttt{vas\_attach} associates it with the current process; a
given VAS may be shared among processes, serving for inter-process
communication, if they all attach to the same VAS. Conversely, a single process
may be associated with several VASes and switch among them programmatically by
calling the \texttt{vas\_switch} function either to perform specific operations
with better isolation or to fluidly manage multiple execution flows.

\subsection{Memory Access Control and Translation}

Making memory available and usable for user applications represents one of the
core OS duties, and most OSes offer the illusion of full memory availability by
decoupling the physical memory from how processes see it. Processes only see a
Virtual Address Space (VAS), which is mapped by the OS to the physical memory,
guaranteeing good isolation among processes. To handle VASes and offer useful
features for user applications, OSes have to adopt a specific memory model;
currently, most of the production-oriented OSes and hardware broadly support
the page-based memory management model. This model divides the VAS and the
physical address space of each process into a set of pages, which is a small
range of contiguous addresses with a fixed size, start address, and
permissions. The page memory model has some attractive advantages: permission
control at the page size level, mechanisms for data sharing, fast protection
checking, accurate notifications about protection violations, and the
possibility of mapping memory to disk.

The approach of using a single address space per process has proven efficient
over the years but, despite its success, it is not flawless and is still open
for improvements. First, the single linear address space approach isolates each
process in memory, which enhances system reliability and security.
Nevertheless, the process has virtually no way to restrict its own access to
some of its memory segments, which might be useful to reduce the security risk
of using third-party binary code. Second, control of memory sharing is limited
to the page size; this makes data sharing less efficient and reduces the
programming possibilities in user space. Finally, the coarseness of page-level
protection creates opportunities for malicious exploits, such as buffer and
stack overflows or code execution in shared libraries.

Motivated by the goal of providing fine-grained control over memory, Witchel et
al. proposed the addition of new hardware features and the development of the
corresponding software abstractions. This approach, called Mondriaan Memory
Protection (MMP)~\cite{mondrian:2002}, enables data access control at the word
size level by inspecting each load/store instruction made by a process to verify
read/write, ownership, and inter-process memory access. To minimize the overhead
of such checks, the authors proposed to extend processor architectures, adding
this permission control at the hardware level. At its core, the MMP implementation
resembles the TLB mechanism of current hardware: a register named Permission
Table Entry (similar to the Page Table Entry) is responsible for keeping a
reference to the Permission Table (similar to the Page Table). At the OS level,
MMP adds data structures that hold the permission information of each process
and a new subsystem named Memory Supervisor, which is responsible for enforcing
the policy and for maintaining low-level data structures. The authors
demonstrated the concept by using simulated hardware and a customized version
of GNU/Linux.

While the mechanism has broad applications, their experimentation with that
approach focused on showing the improvements in system reliability brought by
the isolation of modules in kernel space, a benefit that may extend to dynamic
loaded plugins in user space. The main limitation of this approach is
the dependency on new hardware. In contrast, Swift et al. propose
Nooks~\cite{swift:2003}, a software-only system that implements memory access control
to more thoroughly isolate the kernel from its extensions (modules), improving
OS reliability. Since it cannot make use of hardware-based access control, Nooks
is a best-effort system: it tries to handle programming errors and provide recovery
mechanisms during runtime, but it cannot protect against malicious code nor
every possible coding mistake.

Nooks has two isolation mechanisms: Lightweight Kernel Protection Domain (LKPD)
and Extension Procedure Call (XPC). LKPD is an execution context with kernel
privileges, but with write permission limited to its memory region; whenever a
kernel module is loaded, it is encapsulated in a new LKPD. XPC is the mechanism
used to mediate communication among LKPDs (including the kernel itself). A
function call originating in an LKPD context aimed at a different LKPD context
is first handed to XPC, which performs the necessary checks before finishing
the call (XPC has semantics similar to that of Remote Procedure Calls).
Together, these two mechanisms reduce the risk that memory faults in a given
module (in an LKPD) propagate to other areas of kernel memory.

To implement Nooks in a production-oriented OS, a one-time effort to modify the
kernel functions that interact with extensions in order to add support for XPC
is necessary. Since this does not impose changes to their API, few or no
changes to the modules are necessary. The exception are extensions that export
data structures; Nooks tracks all data structures used in the communication
between the kernel and its extensions to handle this particular scenario.

While Nooks is mostly concerned with kernel-level code, it is relevant to
process abstraction because it also implements a user space recovery mechanism.
Typically, a kernel-level failure causes any process interacting with the
failed module to crash. In Nooks, the system may communicate with the
application, which in turn directs Nooks on how to proceed. For example, the
application may request Nooks to reload the failed module with different
parameters and retry the failed operation, or it may abandon the operation
altogether and pursue an alternative execution path.

\subsection{Hardware Access Control}

The process abstraction is the focal point in OS design, mapping other
abstractions to it; hence, other components of the OS exist to provide the
required mechanisms for orchestrating all processes operations (e.g.,
schedulers and memory management). All needed structures for managing processes
have the side effect of utilizing CPU (overhead); to try to mitigate this
situation, OSes employ a vast number of hardware and software optimizations.

Changes in the process abstraction commonly have impacts on performance, and
the consequences vary according to the proposal. For example, an additional
verification layer can raise the system overhead due to the new feature.
However, while process extensions may degrade performance, they might enhance
overall system performance by exploiting modern hardware features.

An obvious mechanism to improve performance is to shorten the distance between
kernel and user space, providing lower-level access to the application. Engler
et al. introduced a new OS design named exokernel~\cite{exokernel:1995}, famous
for its bold decision of completely removing all abstractions from the OS core,
including processes, and managing hardware access via an OS library. One
advantage of this approach is the offer of multiple different abstractions for
each resource, allowing the application to select the best one for a given task.
This means processes in exokernel can be deeply customized: for example, it is
possible for processes with and without VASes to coexist.

The exokernel approach represents a radical innovation in OS design, especially
in niche applications, but has huge obstacles for adoption by general purpose
OSes. Belay et al. proposed a less radical approach named Dune~\cite{dune:2012},
which brings performance improvements by exploiting hardware virtualization
features in the Linux Kernel. They used Intel VT-x~\cite{intelvt:2005} technology
to provide direct, safe, and secure application access to low-level processor
features such as exceptions, virtual memory, privilege modes, and segmentation.
The authors experimented with these mechanisms by implementing three different
types of application: a sandbox for untrusted code, a privilege separation
facility, and a garbage collector. They reported simplified development and
significant performance improvements. Dune makes few changes in the OS and
provides a straightforward usage mechanism with as little impact as possible
to user applications.

\subsection{Resource Management}

Every application running in the OS consumes system resources. Often, they
perform most of the work in user space, requiring little or no intervention of
the OS. For example, an application that performs complex calculations does not
need much OS intervention. Nonetheless, there is software that demands
significant OS participation to fulfill their goals, extending its consumption
of system resources to the kernel. For example, a network application has part
of its activity conducted by the OS on its behalf when a packet arrives. This
situation may generate problems due to the indirect and uncontrolled use of
resources by activities entrusted to the kernel. Denial-of-service attacks
represent a real-life example of the unrestrained rise in resource consumption
at the OS level.

Banga et al. proposed an OS abstraction named Resource Container (RC)~\cite{banga:1999},
which manages resource consumption by applications bound to it and exports
resource information for both the applications and the scheduler; the latter
can use this information to adapt its algorithm. Processes are bound to a given
RC at startup but may switch to a different one during execution. To manipulate
the RC, the application has an API that defines operations for container
creation, release, and adjustment, thread binding, socket and file binding, and
others.

\section{Potential and Difficulties for Adoption}

Operating Systems researchers produce a vast set of innovations with the
intention of pushing forward the boundaries of the field; however,
production-oriented OSes and research projects have different constraints. The
implementation of new proposals that expand the process abstraction has to
address issues related to compatibility, better use of current hardware
resources, reliability, and be general enough to support multiple programming
languages.

Production-oriented OSes demand strong validation to maintain system
reliability at a variety of scalable configurations: preventing illegal memory
access, API violations, excessive resource consumption, and synchronization or
locking errors are features taken for granted by OS users~\cite{mondrix:2005}.
This makes it hard to adopt research proposals no matter how well they solve
any single specific aspect.

There is a significant set of existing user applications that perform essential
tasks; for example, both web servers and browsers are vital players in the
Internet context. For a new OS feature to be adopted, it is important to
guarantee that such applications will not suffer in performance or ease of use.

Hardware manufacturers continuously develop new features that, once implemented
in servers or niche devices, swiftly spread among end users. An example of this
fast evolution is hardware virtualization: once a server-only capability, it is
now available on most computers. Such new hardware features present additional
unintended opportunities to improve the process abstraction. Nonetheless, such
proposals may have problems related to the dependency on some specific features
which may not be available for all users. For this reason, any change to
processes that require specialized hardware must handle all sorts of corner
cases. Conversely, proposals for improvements in the process abstraction that
suggest changes to hardware could be helpful in pushing chip design forward. Of
course, hardware evolution should take care not to break binary compatibility
with legacy applications. Unfortunately, this may make the ample adoption of
some ideas impractical.

Some of the new process abstraction proposals have dependencies on other
innovative technologies. While this can bring advantages for both the new
process concept and the related technology, it also reduces the chances of the
new abstraction to get adopted in production-oriented OSes due to this
dependence on another potentially unstable technology.

A new proposal of change to the process abstraction has to carefully analyze
the mentioned tradeoffs to achieve production quality. Academia and industry
have to find an equilibrium between research and development to bring benefits
for end users in a timely fashion.

\section{Towards the Next Generation Process Abstraction}

As we examined the works on this field, we recognized a general pattern: most
research on the process abstraction attempts to reduce coupling in one or more
of its elements. It is possible to find radical proposals such as exokernel
that completely decouple the entire OS abstractions; however, most of the
studies adopt a less drastic approach, such as lwC, SpaceJMP, Wedge, and
Resource Containers.

Accordingly, we believe that the \textit{decoupling of VAS, memory isolation,
execution state, privilege separation, resource management}, and maybe others
could bring substantial improvements to production-oriented OSes in terms of
performance, features, security, and simplicity required by next generation
hardware and applications. For example, decoupling such abstractions can become
an excellent option for sharing data and providing integration for
microservices. However, the challenges to mature these ideas are equally
enormous. As a first step to this end, current OSes should move towards
reducing the excessive dependence between its internal elements (e.g., the
Linux process abstraction currently depends on memory management and on the
scheduling interface). This rework in the OS internal components would
facilitate embracing the techniques suggested by different researchers.

\begin{figure*}
    \includegraphics[width=\textwidth]{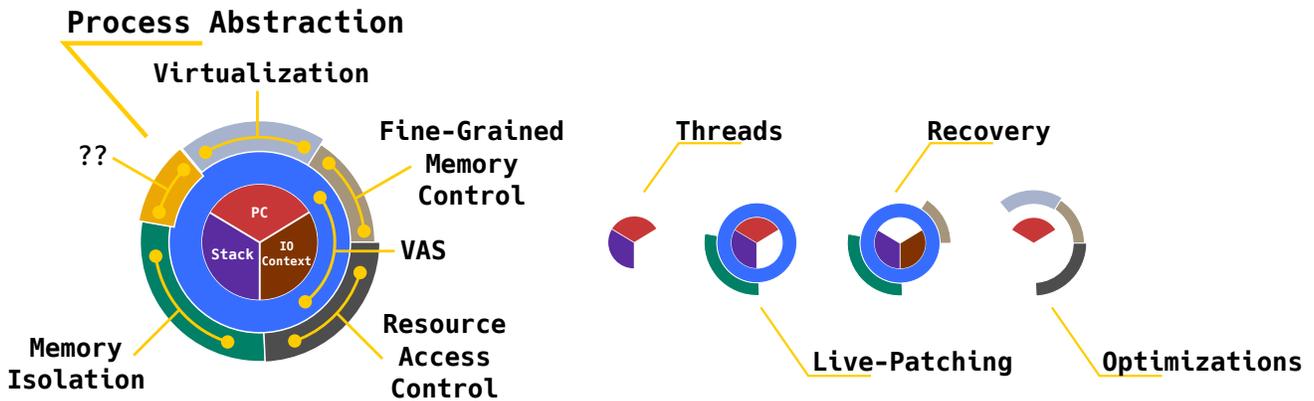}
    \caption{Process Decoupling\label{fig1}}
\end{figure*}

While most research proposals implement a single new specific feature and
rarely integrate with previous work, it is possible to merge key traits of each
one into a unified model, pointing to a new and improved process abstraction.
Figure \ref{fig1} provides a view of a process abstraction with different
elements loosely coupled. Note that some decoupling has already happened; for
example, threads were created by decoupling the Program Counter and the Stack
from the rest of the process components. The figure illustrates a combined view
of the many previously discussed decomposed elements that may form a new
process abstraction. Drawing from the aspects highlighted in Section “Potential
and Difficulties for Adoption”, we now discuss how their adoption might offer
benefits in at least one relevant area.

\subsection{System Reliability and Security Layers}

Industry and Academia agree on the importance of improving system security at
many different levels; as a result, researchers and engineers continuously work
to refine OS security layers. From the process abstraction perspective, such
endeavors focus on enhancing memory access control. Works such as Mondrix,
Nooks, and Wedge have two significant features in common: they all create a new
security level and propose operations to manage the interaction between layers
with different access permissions, controlling the shift in the execution flow
from one layer to the other. Among the proposals discussed, the approach
implemented by Nooks is perhaps the most interesting because it has a
well-defined protocol based on a model that resembles the well-known RPC
pattern. How to create this communication system together with memory isolation
in a production-oriented OS is an engineering and research challenge for the
next generation of the process abstraction.

System reliability is a constant challenge, particularly in user space, and it
is desirable to keep critical applications working for a long time without
human interference. The decoupling of resources may bring benefits in this
regard, by enabling more secure and efficient forms of data sharing, creating
recovery mechanisms in the face of failures, and improving load balancer
algorithms. Nooks, for example, promotes advances in this area by detecting
problems and handling them in tandem with the application.

\subsection{Performance}

The demand for ever higher performance is constant. We should not rely only on
hardware improvements to supply such performance; new software mechanisms are
required. This, in turn, puts additional constraints on the development of
better isolation mechanisms, which should impose minimal performance overhead.
Decoupling some elements of the process abstraction could bring new programming
APIs and features that can deliver these performance improvements.

One major common characteristic of many current and future applications is the
need for parallel or distributed large-scale computing. Memory management
decoupling, as seen in Mondriaan and SpaceJMP, allows for word-sized data
sharing with minimal data copying and reduces interprocess communication
overhead by using shared memory segments with unified pointer addresses.

Areas such as microservices and machine learning can benefit from software
mechanisms for fast initialization and efficient process migration. Decoupling
the VAS, as in SpaceJMP, and the process state (PC), as in \textit{lw}C, makes
it possible to copy the global state of a given application right after startup
and directly load it at initialization in future executions instead of
proceeding with the full initialization routine every time. If the application
is deployed inside replicable containers, their life cycle management may
become significantly faster.

Finally, the shift of virtualization techniques to the process level, as in
Dune, enables performance gains in userspace, by allowing direct lower-level
hardware access, while improving security. This, in turn, obviates the need for
convoluted userspace code aimed at improving performance.

\subsection{Hardware Support}

New hardware trends represent another aspect that processes have to be adapted
for in order to provide new capabilities to user space; additionally, new
hardware could change the way OSes are designed. Fine-grained hardware-assisted
memory access control, as proposed by Mondrix, may boost security in shared or
limited trust environments such as cloud computing. In a different vein, the
current process abstraction is unprepared to handle large-scale, non-volatile,
distributed memory. SpaceJMP tried to anticipate a solution for this problem by
decoupling the VAS and using persistent, shared memory segments.

\subsection{Support for Modernizing Applications}

As previously mentioned, there is a large number of monolithic and legacy
applications that should be updated if they are to fit new paradigms or provide
increased modularization. However, current OSes provide little support to
simplify this task. The combination of live-patching in user space and low-cost
compartmentalization, both made possible by \textit{lw}C, SpaceJMP, Wedge, and
Mondrix, emerges as an alternative to modernize such applications while keeping
backward compatibility. Live-patching in user space could be enabled by the
combination of multiple VASes and fine-grained control over the PC, but that
adds security concerns; to address these, low-cost compartmentalization offers
the necessary fine-grained memory access control. Compartmentalization
techniques may also facilitate the migration of legacy applications to
microservices.

\begin{table*}
\bgroup
\centering
\def\ourStrut{\rule[-.75\baselineskip]{0pt}{2.1\baselineskip}}
\def\ok{\bgroup\color{darkgray}\ding{52}\egroup}
\def\ook{\ok\ok}
\def\oook{\ok\ook}
\sffamily
\fontsize{9.2pt}{11pt}\selectfont
  \begin{tabular}
      {|>{\columncolor{Goldenrod!50!yellow!50!white}\centering}m{5.6em}
       |>{\centering}m{5.7em}
       |>{\centering}m{4.7em}
       |>{\centering}m{5.3em}
       |>{\centering}m{6em}
       |>{\centering}m{5.3em}
       |>{\centering}m{5.2em}
       |
      }
    \hline
    \rowcolor{ForestGreen!50!white}
      \cellcolor{white}Benefits / Decoupling strategy &
      New Programming models &
      Process Persistence &
      Fine-grained privileges control &
      Security Improvements &
      Recovery Mechanisms &
      Performance \tabularnewline
    \hline\ourStrut
                     PC & \oook & \ok  &       &      &       & \ook  \tabularnewline
    \hline\ourStrut
                    VAS & \ook  & \ook &       &      & \ook  &       \tabularnewline
    \hline
    Resource Management & \ok   &      &       &      & \oook & \ok   \tabularnewline
    \hline
       Memory Isolation &       &      & \oook & \ook &       &       \tabularnewline
    \hline\ourStrut
         Virtualization &       &      &       & \ok  &       & \oook \tabularnewline
    \hline\ourStrut
             Privileges & \ook  &      & \oook & \ook &       &       \tabularnewline
    \hline
  \end{tabular}
  \caption{Property and Potential Improvements. We use a range of zero
  to three checks to highlight the relevance of each process abstraction
  advancement for each user space application.\label{tab1}}
\egroup
\end{table*}

\section{Future Directions}

The characteristics and challenges presented throughout this paper, in our
view, outline the research opportunities for industry and academia to push
forward production-oriented OSes design. These improvements may bring security
benefits, provide new user space features, offer optimizations, among countless
other possibilities both to future and legacy applications.

In general, the evolution of the process abstraction should go towards the
decoupling of several components; such decoupling could open new and exciting
opportunities for OS design and user space applications, as summarized on
Table~\ref{tab1}.


\printbibliography

\setlength\parindent{0pt}

\newpage
\section*{About the Authors}

\textbf{Rodrigo Siqueira} is a Linux kernel developer at Advanced Micro Devices (AMD) and has a Masters's degree in Computer Science from the Institute of Mathematics and Statistics of the University of São Paulo (IME-USP). His research interests include Operating System, GPU,  Software Engineering, and Free Software. Additionally, he contributes to free software communities, such as Linux Kernel and Debian.

\textbf{Nelson Lago} has a Masters degree in Computer Science and is the technical manager for the CCSL at IME-USP, where he regularly participates in public debates on issues such as software patents, privacy, and copyright. His research interests gravitate around Free Software, Computer Music, and Distributed Systems.

\textbf{Fabio Kon} is a Full Professor of Computer Science at the University of São Paulo and Special Advisor to the Scientific Director at the S\~ao Paulo Research Agency. His research interests gravitate around the design, implementation, and assessment of Complex Software Systems.

\textbf{Dejan Milojičić} is a distinguished technologist at Hewlett Packard Labs. His research interests include OSes, distributed systems, and systems management. Milojičić received a PhD from University of Kaiserslautern.

\end{document}